%% file: ms.tex
\documentclass[12pt,preprint]{aastex}

\shorttitle{NEP Survey: X-ray Data}
\shortauthors{Henry et al.}

\begin{document}

\title{The {\it ROSAT} North Ecliptic Pole Survey: The X-ray Catalog}

\author{J. Patrick Henry\altaffilmark{1}}
\affil{Institute for Astronomy, University of Hawai'i, 2680 Woodlawn
Drive, Honolulu, HI 96822; \\ and Max-Planck-Institute f\"ur
extraterrestrische Physik, Giessenbachstrasse, Postfach 1312,
Garching, D-85741, Germany}
\email{henry@ifa.hawaii.edu}

\author{Christopher R. Mullis\altaffilmark{1}}
\affil{Department of Astronomy, University of Michigan, 501 E.
University Avenue, Ann, Arbor, MI 48109}

\author{Wolfgang Voges, Hans B\"ohringer, and Ulrich G. Briel}
\affil{Max-Planck-Institute f\"ur extraterrestrische Physik, 
Giessenbachstrasse, Postfach 1312, Garching, D-85741, Germany}

\author{Isabella M. Gioia\altaffilmark{1}} \affil{Istituto di
Radioastronomia INAF - CNR, Via Gobetti 101, I-40129, Bologna, Italy;
\\ and Institute for Astronomy, University of Hawai'i, 2680 Woodlawn
Drive, Honolulu, HI 96822}

\and

\author{John P. Huchra\altaffilmark{2}}
\affil{Harvard-Smithsonian Center for Astrophysics, 60 Garden Street,
Cambridge, MA 01970}

\altaffiltext{1}{Visiting Astronomer at the Canada-France-Hawai'i
Telescope, operated by the National Research Council of Canada, le
Centre National de la Recherche Scientifique de France and the
University of Hawai'i, at the W. M. Keck Observatory, jointly operated
by the California Institute of Technology, the University of 
California, and the National Aeronautics and Space Administration and
Observer at the University of Hawai'i 2.2m telescope.}
\altaffiltext{2}{Some observations reported here were made at the
Multiple Mirror Telescope Observatory, a joint facility of the
Smithsonian Institution and the University of Arizona.}

\begin{abstract}
The sky around the North Ecliptic Pole (NEP), at $\alpha$(2000) =
18$^h$00$^m$00$^s$, $\delta$(2000) = +66\degr33\arcmin39\arcsec, has
the deepest exposure of the entire {\it ROSAT} All - Sky Survey
(RASS). The NEP is an undistinguished region of moderate Galactic
latitude, $b=29\fdg8$, and hence suitable for compiling statistical
samples of both galactic and extragalactic objects. We have made such
a compilation in the 80.6 deg$^2$ region surrounding the NEP. Our
sample fully exploits the properties of the RASS, since the only
criteria for inclusion are source position and significance, and
yields the deepest large solid angle contiguous sample of X-ray
sources to date. We find 442 unique sources above a flux limit
$\mathrm{\sim2\times10^{-14} ~ergs ~cm^{-2} ~s^{-1}}$ in the 0.5--2.0
keV band. In this paper we present the X-ray properties of these
sources as determined from the RASS. These include positions, fluxes,
spectral information in the form of hardness ratios, and angular
sizes. Since we have performed a comprehensive optical identification
program we also present the average X-ray properties of classes of
objects typical of the X-ray sky at these flux levels. We discuss the
use of the RASS to find clusters of galaxies based on their X-ray
properties alone.
\end{abstract}

\keywords{galaxies: active --- galaxies: clusters: general --- 
surveys --- X-rays: galaxies: clusters --- X-rays: general ---
X-rays: stars}

\section{INTRODUCTION}

The {\it ROSAT} All - Sky Survey (RASS) \citep{tru83, vog99}, the
first X-ray survey of the sky performed with an imaging detector, is
the X-ray analog of the Palomar Observatory Sky Survey (POSS)
\citep{rei91}. Unlike the POSS the depth of the RASS varies by a
factor of $\sim$10 over the sky. The deepest regions are at the north
and south ecliptic poles, due to the satellite scan restriction that
comes from keeping the solar panels pointed at the Sun.  The South
Atlantic Anomaly reduces the exposure at the South Ecliptic Pole, and
the Large Magellanic Cloud obscures the extragalactic sky there. The
North Ecliptic Pole (NEP) attains the deepest exposure in the RASS
while being an undistinguished spot of moderate galactic latitude and
extinction.

For these reasons, the NEP is the prime target for a deep, contiguous
survey conducted with the {\it ROSAT} satellite. There are other
contiguous {\it ROSAT} surveys, which cover about 100 times more solid
angle than the high exposure region around the NEP but are typically a
factor of 10 shallower. Among these are the BCS and its extension the
eBCS \citep{ebe98, ebe00}, the RASS1-BS \citep{deg99}, the NORAS
\citep{boh00}, the RBS \citep{sch00}, the MACS \citep{ebe01} and the
REFLEX \citep{boh04}. All of these surveys except the RBS are aimed at
compiling X-ray selected cluster catalogs while the first goal of the
NEP survey is to compile a catalog of all X-ray sources in the survey
region. The unique combination of depth plus wide, contiguous sky
coverage provides the capability of both detecting high-redshift
objects and studying large-scale structure, which were the second and
third goals of the survey.

We have previously summarized the main results from the NEP
Survey. \citet{hen01} give an overview of the survey;
\citet{vog01} summarize the X-ray data and source statistical
properties; \citet{gio01} give evidence for cluster X-ray luminosity
evolution; \citet{mul01} describe the properties of the NEP
supercluster. This paper is one in a series that give the complete
details of the survey. \citet{gio03} present the optical
identification program, including the methodology used, and give the
optical properties of all the sources. We note that the NEP survey
features a very high identification rate (99.6\%). \citet{mul04} give
the AGN correlation function measured using the NEP sample, one of the
first such measurements at low redshift. They also give a revised
catalog of AGN sources. 

In this paper we give the X-ray properties of the NEP sources that can
be determined from the RASS. These include position, flux, hardness
ratios and angular sizes. We also give the source selection function
(sky coverage) in one and two dimensional form.  These selection
functions are needed to exploit the complete nature of the catalog. In
addition we summarize the optical properties of each source, including
the identification and redshift if it is extragalactic. A
comprehensive description of the NEP Survey and its principal results
may by found in \citet{mul011}.

\section{OBSERVATIONS}

{\it ROSAT} was launched on 1990 June 1 and the all-sky survey was
conducted from 1990 July 30 to 1991 January 25 with an additional
three ``fill-in'' periods of three to ten days duration.  The {\it
ROSAT} survey strategy yielded the longest exposure at the NEP, which
was more than one hundred times longer than the average exposure.  The
equatorial coordinates of the NEP are $\alpha$(2000) =
18$^h$00$^m$00$^s$, $\delta$(2000) = +66\degr33\arcmin39\arcsec, its
galactic coordinates are $l$ = 96\fdg4 , $b$ = 29\fdg8; its
supergalactic coordinates are $l_{SG}$ = 33\fdg4, $b_{SG}$ = 38\fdg3;
and its neutral hydrogen column density is $\mathrm{4.1\times10^{20}
~cm^{-2}}$ \citep{elv94}.

We used the second processing of the observations (RASS-II), performed
in 1994 to 1995, for our work \citep{vog99}. During this processing
photons were merged into 1,378 6\fdg4 $\times$ 6\fdg4 sky fields that
overlapped by at least 0\fdg23 so objects would not be lost at sky
field boundaries during the source detection phase of the
analysis. The layout of these fields on the sky can be found at
$http://www.xray.mpe.mpg.de/rosat/survey/rass-3/sup/psplit.ps.gz$.

In addition, we used an improved aspect solution relative to RASS-I
and rejected photons acquired during times when the solution was
poor. Figure 1 shows the equivalent on-axis exposure achieved as a
result of the new aspect solution. This exposure map is produced by
scanning the detector map over the sky in the same pattern as the
survey.  As explained in \citet{sno94} the detector map includes the
effects of vignetting and the detector window support structure, all
normalized to unity at the center of the field of view. The absolute
value of the transmission of the window plus its support structure is
incorporated into the detector response matrix, see Section 2.3. The
minimum, median, and maximum equivalent on-axis exposure for the NEP
9\degr $\times$ 9\degr~region are 1.7, 4.8, and 40 ks, respectively.
The 2\degr~diameter peaked region surrounding the NEP is the field of
view of the PSPC. The exact NEP is a local exposure minimum due to the
wagon wheel - shaped detector window support structure.

The point spread function (PSF) is an important ingredient of the
survey analysis. It is used not only to separate point-like from
extended sources, but also to correct for flux outside of the source
detection region. \citet{boe00} describes this function for the RASS,
which is the sum of off-axis PSFs vignetting weighted according to the
scan pattern of the survey. This calculation makes the assumption that
the angle between successive scan lines is small ($\leq4$\arcmin),
which is not valid within 1\degr~ of the NEP. Nevertheless we adopt it
for the entire NEP Survey since the affected solid angle is only 4\%
of the total solid angle of the survey. The RASS PSF half power
diameter is 2\farcm8.  For computational reasons the PSF is further
approximated as a Gaussian by the analysis system when discriminating
between extended and point sources and measuring source
size. \citet{boe00} describes this approximation as well.

\subsection{Souce Detection}

A detailed description of source detection in the RASS is given by
\citet{vog99}. We review the procedure here, since it forms the basis
of the entire NEP Survey, and also provide some details particularly
relevant to our survey. We report count rates in the broad band and
fluxes and luminosities in the 0.5 - 2.0 keV band.

Sources are detected in a three-stage process. The first two stages
use sliding boxes operating on binned RASS photons sorted into the sky 
fields comprised of $512 \times 512$ $45\arcsec$ pixels. Allowing for 
overlaps, 8 such sky fields are required for the NEP Survey region.

In the first stage a LDETECT, or local detection, algorithm is used
separately for soft, (0.1 - 0.4 keV or PHA channels 11 - 41), hard
(0.5 - 2.0 keV or PHA channels 52 - 201) and broad (0.1 - 2.4 keV or
PHA channels 11 - 235) band photons. This method employs a local
measurement of the background. The initial box size is $2\farcm25
\times 2\farcm25$ ($3 \times 3$ pixels). The local background comes
from an annulus $0\farcm75$, or 1 pixel, wide surrounding the
box. Hence the ratio of background to source pixels is 16 to
9. Extended sources are detected by repeating the LDETECT four more
times, doubling the box size each time while maintaining the
background annulus to source areas in the 16 to 9 ratio. The detection
threshold is purposefully set low, likelihood $\ge6$, in order not to
miss real sources.

We detemine the background in the three bands for each sky field using
the following procedure. Circular regions that have similar sizes as
the detect box around the LDETECT sources are excised from the sky
fields. The remaining data are rebinned into coarse pixels, fitted by
a two-dimensional spline, coarse pixels more than $4 \sigma$ above the
spline fit removed, and the whole process iterated until no more
coarse pixels are removed.  Figure 2 shows the resulting background
map in the broad band for the NEP region.

The second stage uses a MDETECT, or map detection, algorithm on the
three bands. The same multi-pass multi-scale box process is used as
with LDETECT with the background coming from the map. Sources with
likelihood $\ge6$ are again retained.

The third and last stage uses a maximum-likelihood (ML) algorithm
\citep{chs88, bd01} that both detects the sources and characterizes
them.  This stage uses the unbinned photons associated with the merged
unique sources found at the LDETECT and MDETECT stages. The ML fit
weights each photon with the {\it ROSAT} mirror + PSPC PSF appropriate
to the energy and off-axis angle at which it was detected. Higher
weight is given to photons received when the source is near the center
of the field rather to those when it is near the edge since the PSF is
a strong function of off-axis angle. Higher weight is also given to
times of lower background rate.

The ML analysis yields a number of source parameters: the likelihood
that the source exists, the likelihood that the source is extended,
its position, fractional error on the net detected counts, and angular
extent. All these parameters are for the broad energy band. The
likelihood that a source exists (Exist L or L for short) is L = -lnP
where P is the ML probability that the source count rate is
zero. \citet{bd01} provide a detailed theoretical treatment of this
problem.

After the source is significantly detected and its position
determined, its count rate in the broad band is calculated using
circular aperture photometry with a $5\arcmin$ radius (for all but RX
J1834.1+7057) centered on its position. The circle is divided into an
inner $2\farcm5$ radius region and eight equal-sized sectors in the
remaining $2\farcm5$~- $5\arcmin$~annulus. Any sector containing
another significant source is discarded. The counts in the central
region and remaining sectors are summed. Net counts result from
subtracting the map background. Dividing by the on-axis exposure in
Figure 1 gives the net vignetting-corrected counting rate. All count
rates in this paper (except RX J1834.1+7057) are those derived by this
procedure. We refer to the corresponding fluxes or luminosities as
detect fluxes or luminosities. The minimum and maximum count rates are
(1.2$\pm$0.2) $\times~ 10^{-3}$ cts s$^{-1}$ and 1.07$\pm$0.01 cts
s$^{-1}$. The median fractional error on the count rate is 16\%.

In addition to the above analysis in the broad band, an existence
likelihood and count rate using the same procedure, but with the
position fixed to that found for the broad band, are determined in
four energy bands. The first two are the soft and hard bands described
above. The third and fourth are called C and D and are the 0.52 - 0.90
keV (PHA channels 52 - 90) and 0.91 - 2.01 keV (PHA channels 91 - 201)
ranges respectively. Hardness ratios comparing the hard to soft bands
(HR1) and the D to C bands (HR2) are then calculated.

\subsection{Selection Criteria and Functions}

We give the selection criteria for inclusion in the {\it ROSAT} NEP
Survey in Table 1. There are 442 uniques sources that meet these
criteria. There were twenty multiple detections of the same, often
extended, source. All but one was deleted after a visual
inspection. The only exception to this deletion step was for RX
J1724.1+7000 (NEP 1590) and RX J1724.2 +6956 (NEP 1591), which we
classify as two pieces of a single source but retain both in our final
catalog. The reason for doing so is because there may be another
cluster with redshift 0.23 behind the southern source, based on three
concordant redshifts. There are 518 additional sources in the survey
region that do not meet the existence likelihood criterion and are
therefore not in the sample. Of these only 3 have a detect count rate
signal-to-noise ratio $>4\sigma$. Thus the fundamental selection
criterion for inclusion in the {\it ROSAT} NEP Survey is the detect
count rate threshold. Figure 3 shows the source distribution on the
sky and the source class.

The survey selection function or sky coverage is the solid angle in
which sources of a given count rate could have been detected. This
function is needed to transform the list of sources given here into
quantitative statistical statements, such as logN - logS relations,
spatial correlation functions or luminosity functions.

We model the complicated detection process described above with an
analytic signal-to-noise calculation involving a circular detect cell.
Background is an important component of this calculation because it is
more than a factor of ten higher than the average for the RASS. We may
ignore the error on the background because it comes from a map
determined over a 6\fdg4 $\times$ 6\fdg4 sky field, i.e. the error on
the source counts is much larger, coming from a 5\arcmin ~radius
circle. In this situation the signal-to-noise ratio on the source flux
is.
\begin{equation}
\frac{S}{N} = \frac{R_{S} T}{\sqrt{R_{S} T+R_{B} T \pi r_{d}^{2}}},
\end{equation}
where $R_{S}$ is the source count rate in the 5\arcmin ~aperture,
$R_{B}$ is the background counting rate per square arcminute, $T$ is
the exposure time and $r_{d}$ is the effective detect cell radius in
arcminutes.

We must first determine r$_{d}$, which calibrates the simple photon
statistics of our model to the actual maximum-likelihood procedure.
The maximum-likelihood detection method is more sensitive than the
sliding circle method because the former incorporates the
instantaneous PSF and background when each photon was detected. We
thus expect $r_{d}$ will be smaller than the 5\arcmin ~radius aperture
used to determine the source count rate. We selected 71 sources from
the NEP region with maximum likelihood count rates measured between
3.8 $\sigma$ and 4.2 $\sigma$. The best agreement between the measured
S/N and that predicted from equation (1) with the actual $R_{S}$ and
$T$ (from Figure 1) and $R_{B}$ (from Figure 2) at the position of the
sources is with r$_{d}$ = 1\farcm58. Figure 1 of \citet{hen01} shows
the comparison. This value is the $\sim55$\% encircled energy radius
of the survey PSF \citep{boe00}.

Applying equation (1) with S/N = 4 and $r_{d}$ = 1\farcm58 to an array
of 720 $\times$ 720 points in the exposure and background maps of
Figures 1 and 2 yields a map of the count rate $R_{S}$ that can be
measured at $4\sigma$ confidence. This map is the fundamental 2D
survey selection function and we show it in Figure 4.  We integrate
this map to yield the total solid angle surveyed as a function of
count rate or 1D survey selection function given in Figure 5 and
Table 2. 

The total solid angle is 80.6 deg$^{2}$ at high fluxes. The average
source density on the sky is 5.5 deg$^{-2}$. The count rate limit in
the 5\arcmin ~radius flux aperture and in the 0.1 - 2.4 keV band at
half that solid angle is 0.0074 ct s$^{-1}$. Under the assumptions
described in Sections 2.3 and 2.4 this rate comes from an
extragalactic point source with a total unabsorbed flux of $7.5 \times
10^{-14}$ erg cm$^{-2}$ s$^{-1}$, or a galactic source with a total
unabsorbed flux of $4.6 \times 10^{-14}$ erg cm$^{-2}$ s$^{-1}$ or a
cluster with an unabsorbed surface brightness averaged over a
$5\arcmin$~radius aperture of $1.0 \times 10^{-15}$ erg cm$^{-2}$
s$^{-1}$ arcmin$^{-2}$, all in the 0.5 - 2.0 keV band. There are two
sources fainter than the survey limit of 0.002 ct s$^{-1}$. It is
likely that we did not sample the survey area densely enough when
computing the sky coverage to recover the very small region of sky
probed deeper than this limit.

\subsection{Counts to Flux Conversion}

We next need to convert the ROSAT PSPC count rates into fluxes in
order to compare with others and to derive various physical parameters
associated with the sources. We determine the conversion by folding
model spectra through the response matrix (pspcc\_gain1\_256.rsp) to
yield a counting rate in PHA channels 11 - 235. The response matrix is
a convolution in photon energy space of the detector redistribution
matrix with the transmission of the detector window (including the
transmission of its support structure) times the effective area of the
telescope.  An appropriate integration of the model spectrum yields
its flux in 0.5 - 2.0 keV band that we use. The ratio of these two
quantities gives the energy conversion factor (ECF).

Were it not for the variable extinction across the survey region, we
could derive a unique ECF for each source type. However, the neutral
Hydrogen column densities in the NEP region vary from a minimum of
$2.5 \times 10^{20}$ cm$^{-2}$ to a maximum of $8.3 \times 10^{20}$
cm$^{-2}$ with a median of $4.1 \times 10^{20}$ cm$^{-2}$
\citep{elv94, str92}.  This range is small, nevertheless we derive a
separate ECF for each source, considering the column density in its
direction. In order to compare sources, we quote unabsorbed fluxes,
that is those if there were no Milky Way absorbing material.

We assume different unabsorbed spectra for different classes of
objects. For galactic objects (stars and a single planetary nebula) we
use a Raymond-Smith spectrum \citep{rs77} with a temperature of
$10^{7}$K and solar abundances. Further, we assume no galactic
absorption, hence there is a unique ECF of $5.95 \times 10^{-12}$ erg
(0.5 - 2.0 keV) cm$^{-2}$ ct (0.1 - 2.4 keV)$^{-1}$. Since all
galactic objects in the survey are point sources, the detect fluxes in
the 5\arcmin ~radius flux aperture are multiplied by 1.0498 to obtain
the total flux (see Section 2.4 for a description of this size
correction).

For extragalactic point sources (AGNs, BL Lacs and a single galaxy),
we use a power-law spectrum with photon index of $\Gamma = 2$, where
the photon spectrum is proportional to E$^{-\Gamma}$. The median ECF
is $9.72 \times 10^{-12}$ erg (0.5 - 2.0 keV) cm$^{-2}$ ct (0.1 - 2.4
keV)$^{-1}$ and it varies by $\pm30$\% over the full range of column
desnities. Total fluxes again come from the factor of 1.0498
correction to the detect flux for point sources. We calculate
K-corrected luminosities in the 0.5 - 2.0 keV band in the source frame
for a cosmology with H$_0$ = 70 h$_{70}$ km s$^{-1}$ Mpc$^{-1}$,
$\Omega_{m0}=0.3$ and $\Omega_{\Lambda0}=0.7$, where H$_0$,
$\Omega_{m0}$ and $\Omega_{\Lambda0}$ are the present values of the
Hubble parameter, the matter density relative to critical and the
energy density relative to critical, respectively. Note that the
K-correction for a power-law photon spectrum is $(1+z)^{\Gamma-2}$ or
unity for our assumed spectrum.

Obtaining the ECF for extragalactic diffuse objects (groups and
clusters of galaxies) is considerably more involved because their
spectra depend on their luminosity and they are not point sources. We
use a MEKAL spectrum \citep{kaa92, lie95} with metallicity of 0.3
solar, temperature and redshift particular to each object and
Hydrogen column density in the direction of the object. The redshift
is that measured from our identification program \citep{gio03}. The
temperature is estimated from the low-redshift luminosity-temperature
relation \citep{wjf97}, assumed not to evolve.
\begin{equation}
kT=3.45~keV~[h^{-2}_{70}L_{bol,44}]^{0.33}
\end{equation}
where L$_{bol,44}$ is the bolometric luminosity in units of 10$^{44}$
erg s$^{-1}$ and we have scaled the coefficient to H$_0$ = 70 km
s$^{-1}$ Mpc$^{-1}$. We begin an iterative procedure by assuming a
cluster temperature of 3 keV, combined with the redshift as needed, to
derive an ECF, r$_{200}$ (for the size correction, see Section 2.4), a
K-correction, and a bolometric correction. We then derive the
bolometric luminosity by combining these four quantities, the redshift
and the detect counts, for the cosmological parameters given
above. Insertng the bolometric luminosity into equation (2) yields a
revised temperature and the loop is repeated until the temperature
converges. No more than three iterations are required for the
temperature to converge to within 5\%. We compute the final ECF, size
correction, and K-correction from the final iterated temperature.  The
median cluster ECF is $1.08 \times 10^{-11}$ erg (0.5 - 2.0 keV)
cm$^{-2}$ ct (0.1 - 2.4 keV)$^{-1}$ and it varies by $\pm17$\% over
the full range of column densities, redshifts, and counting rates of
the NEP clusters. Note that low redshift groups can be problematic
since they have large angular sizes (large size corrections) and low
temperature complex spectra dominated by line emission.

\subsection{Size Correction}

The X-ray count rates determined for the NEP survey are detect
rates, those within a circular aperture. The size correction gives the
total rate as R$_{tot}$ = R$_{S}~\times$ (size correction). All point
sources (stars, AGNs, BL Lacs, the planetary nebula and the galaxy)
have aperture sizes of 5\arcmin ~radius and the size correction is the
reciprocal of the integral of the PSF out to this radius or 1.0498 for
the 1 keV PSF of \citet{boe00}.  We apply this correction for
completeness, but note that it is smaller than some systematic
uncertainities, e.g. the effective area. We further note that this
value is very slightly different from that in our previous work
(1.0369), which used a preliminary RASS point spread function.

Groups and clusters of galaxies are extended sources whose apparent
size varies with redshift, of course, and also from object to object
(clusters are not all the same intrinsic size). However there is not
enough information to derive that size from the data, given the small
numbers of detected photons for most sources. Thus our correction must
necessarily be an approximate ``one size fits all'' procedure. The
flux measurement aperture is again 5\arcmin ~for all sources except
RX J1834.1+7057, which has an aperture radius = 6\farcm5, so the
correction is not too large ($<$ 50\%) for all but the closest
clusters, as we show in Figure 6. Conversely, clusters with redshifts
greater than 0.5 are effectively point sources in this context. We
convolve the 1 keV PSF with a beta model surface brightness
distribution to give the observed surface brightness. The parameters
of the beta model are $\beta$ = 2/3 and nonevolving core radius
r$_{c}$ = 180 h$^{-1}_{70}$ kpc, corresponding to the traditional 250
h$^{-1}_{50}$ kpc. The size correction is the integral of the true
surface brightness out to r$_{200}$ (the ``edge'' of the cluster)
divided by the integral of the observed surface brightness out to the
aperture radius (r$_a$ = 5\arcmin ~or 6\farcm5):
\begin{equation}
\frac{r^2_c\{1-[1 + (r_{200}/r_c)^2]^{3/2-3\beta}\}/2/(3\beta-3/2)}
{\int^{r_a}_{0}drr\int^{r_{200}}_{0}dqq[1+(q/r_c)^2]^{1/2-3\beta}
\int^{2\pi}_{0}d\phi PSF(\sqrt{q^2+r^2-2qrcos\phi}}
\end{equation}
We obtain r$_{200}$ by inserting the iterated tempeature into
\begin{equation}
r_{200}=\frac{2.68h^{-1}_{70}Mpc\sqrt{kT/10 keV}}
{\sqrt{\Omega_{m0}~(1+z)^{3}+\Omega_{\Lambda0}}}
\end{equation}
where this relation is scaled from \citet{evr96}. We use the
cosmological parameters given previously to convert from physical to
angular radii and to obtain r$_{200}$. We show the size corrections in
Figure 6. The minimum, median, and maximum values are 1.04, 1.16, and
3.66, respectively. The $<1$\% difference between the minimum size
correction for groups and clusters and that of a point source gives an
estimate of the accuracy of our numerical evaluation of the triple
integral in equation (3). At the median redshift of the NEP clusters,
0.2, the size correction varies only $\pm$2\% for core radii between
140 and 220 h$^{-1}_{70}$ kpc. The value of the corrections and their
variation with assumed core radius are both smaller than those of the
EMSS, for example \citep{gio90, hen92}.

\section{EXAMPLES}

In Table 3 we give some of the X-ray properties of a star, an AGN, and
a cluster of galaxies. These three particular sources are not
representative since they have high signal to noise. However they
introduce the tabular material given in the next section and
illustrate the different properties of each class discussed in Section
5. These three classes represent the bulk of the faint X-ray source
population at the flux levels probed by the NEP Survey.  Figures 7, 8,
and 9 show the Digital Sky Survey scans of the POSS-II red image
overlaid with the RASS X-ray contours \citep{mul011} (see also
$http://www.ifa.hawaii.edu/\sim mullis/nep-catalog.html$). The AGN and
brightest cluster galaxy are visible in the 90\% confidence positional
error circle; the overexposed star obscures its error circle but is
approximately centered on it.

Considering the wider energy range HR1, the data in Table 3 show that
the cluster has a harder spectrum than the AGN, which is harder than
the star. The same trend is apparent for HR2 values of the star and
AGN.

The cluster has a highly significant extent measurement.  The star is
also measured to be extended. Given its optical brightness (V = 4.9),
the stellar identification is very secure and the nonzero measured
extent must be spurious. This effect comes from approximating the
point spread function by a Gaussian, the tails of which are narrower
than the actual point spread function \citep{boe00}. Thus point
sources with very high signal to noise appear extended to the analysis
system We did not use the RASS-derived extent in our source
identification process for this reason.

The AGN has a significantly bluer O - E color than the brightest
cluster galaxy (0.64 vs. 2.24). Faint blue point-like sources within
the 90\% positional error circle are almost always AGN; 68\% of all
AGN in the NEP Survey are bluer than O - E = 1.27 and 90\% are bluer
that O - E = 1.91. These AGN in turn are highly likely to be the
identification given their surface density \citep{boy88}. For example,
at O $<$ 20.7 (approximately B $<$ 20.7) 190 AGN are observed in the
90\% error circles where 1 AGN is expected at random.

\section{THE {\it ROSAT} NORTH ECLIPTIC POLE SURVEY CATALOG OF X-RAY 
\\SOURCES}

We present in Table 4 the basic X-ray properties of the 442 unique
sources that comprise the {\it ROSAT} NEP Survey plus the nature of
the source and its redshift (if extragalactic) from the optical
identification program \citep{gio03}. Each source is listed in order
of increasing right ascension and has two rows in the table. We
describe the content of the columns below.

Column (1) is the X-ray source name and an internal identification
number.

Column (2) is the right ascension and declination of the X-ray
centroid, epoch J2000.

Column (3) is the net count rate and its error in the 0.1-2.4 keV or
PHA channels 11 - 235 (broad) energy band within a circular aperture
of 5\arcmin ~radius, except for RX J1834.1+7057 where the radius is
6\farcm5. The rate is corrected for vignetting, that is it is larger
than the net counts divided by the time the source was in the PSPC
field of view. The error is the 1$\sigma$ uncertainity from the
maximum likelihood procedure described in Section 2.1.

Column (4) is the equivalent on-axis exposure time, (that is, it is
smaller than the time the source was in the field of view) and the
Hydrogen column density from \citet{elv94} supplemented by
\citet{str92}. The column densities were linearly interpolated among
the four pixels of the relevant map nearest to the X-ray position.

Columns (5) and (6) are the hardness ratios 1 and 2 with their
1$\sigma$ uncertainities, respectively. HR1 compares the 0.52 - 2.01
keV band to the 0.11 - 0.41 keV band (PHA channels 52 - 201 and 11 -
41, respectively) while HR2 uses the 0.91 - 2.01 keV and 0.52 - 0.90 keV
bands (PHA channels 91 - 201 and 52 - 90, respectively). The ratios
are calculated dividing the harder minus the softer band net counts by
the harder plus the softer band net counts. Negative net counts
resulting from background subtraction have been set to zero, yielding
hardness ratios between $-1$ and +1.

Column (7) is the sigma of the source extent in arcseconds,
approximating it by a Gaussian, and the difference in likelihoods
between the best fitting extended and point source models of the 
source surface brightness, both measured as described in Section 2.1.

Column (8) contains two source existence likelihoods, from the maximum
likelihood analysis in the broad band and from the MDETECT analysis in
the band for which the source has the highest existence likelihood. In
both cases the value has been set to 999 if it exceeds that value.

Column (9) is the total observed unabsorbed flux in units of
$10^{-14}$ erg cm$^{-2}$ s$^{-1}$, obtained from the detect count rate
in column (3), and the total luminosity in units of $10^{44}$ erg
s$^{-1}$, assuming a cosmology with H$_0$ = 70 km s$^{-1}$ Mpc$^{-1}$,
$\Omega_{m0}=0.3$ and $\Omega_{\Lambda0}=0.7$. Both of the quantities
are in the 0.5 - 2.0 keV energy band, observed and rest, respectively.
Their error may be scaled from that on the count rate. The spectra
assumed are thermal with kT = $10^{7}$~K for galactic objects and that
given in column (11) for galaxy groups and clusters. For extragalactic
point sources, the spectra are power laws with photon indexes
2. Absorption for extragalactic objects is parameterized by the
Hydrogen column density in column (4).

Column (10) is the optical identification class and spectroscopic
redshift for extragalactic objects.  The identification classes are
STAR, CL for galaxy group or cluster, AGN1 or AGN2 based on the
equivalent width of the emission lines and broadness of the permitted
emission lines as defined in \citet{gio03}, BL for BL Lac, PN for
planetary nebula, and GAL for a possibly interacting galaxy.

Column (11) is the parameter of the unabsorbed spectrum, temperature
in keV or photon index, and the size correction. These quantities vary
for groups and clusters but the unabsorbed spectra and size
corrections are the same for all galactic objects and for all
extragalactic point sources, as described in sections 2.3 and 2.4
respectively. In particular, the unabsorbed spectrum for all galactic
sources is a Raymond - Smith thermal plasma model with solar
abundances and a temperature of 10$^7$ K (0.9 keV).  For all
extragalactic point sources the spectrum is a power law with photon
index 2. The size correction is 1.0498 for all point sources.

\subsection{Comparison with Previous NEP Catalogs}

There are a few minor differences between the catalog presented here
and previous NEP catalogs. These differences should be kept in mind
when comparing them. The following four changes have been made with
respect to the optical identification catalog \citep{gio03} and
\citet{mul011}. The assumed cosmology here has H$_0$ = 70 km s$^{-1}$
Mpc$^{-1}$, $\Omega_{m0}=0.3$ and $\Omega_{\Lambda0}=0.7$, instead of
the ``X-ray astronomer's universe'' or 50, 1.0 and 0.0 respectively.
We integrate out to the temperature - dependent r$_{200}$ radius to
obtain the total cluster flux instead of to infinity. The conversion
of counts to flux for AGNs and BL Lacs here assumes a power law
spectrum with photon index of 2 instead of 1. The three sources whose
properties are given in Table 5 (whose columns are identical to those
of Table 4) have been removed from the sample because their existence
likelihood is $\geq 10$ in the soft or hard bands, not in the broad
band. These sources are real sources with firm identifications, but
they do not meet all of the selection criteria specified in Table
1. The count rates, fluxes, luminosities and existence likelihoods
reported previously were for the soft, hard and hard bands (in order
of right ascension).

There is a 4\% standard deviation between the Hydrogen column density
presented here and those in the previous two catalogs and the AGN
catalog in \citet{mul04} due to a rounding error. This change produces
a 1\% standard deviation change in the derived X-ray fluxes and
luminosities, insignificant compared to their 16\% median Poisson
error from photon count statistics.

Finally, the identifications of three sources have been revised. RX
J1806.4+7028 (NEP 4170) is a group not a single galaxy, confirming the
suspicion of \citet{gio03}, because an XMM observation shows the
source is extended (Mullis et al., in preparation). We changed the
classification of RX J1824.7+6509 (NEP 5500) from a star to an AGN1 as
discussed in \citet{mul04} and RX J1724.9+6636 (NEP 1640) from an AGN1
to an AGN2 as discussed in \citet{wol05}.

\section{ENSEMBLE STATISTICAL PROPERTIES OF THE X-RAY SOURCES}

Most NEP sources are detected at threshold, so their individual
properties are only known with large uncertainities. However their
ensemble properties can provide some useful average
characterizations. Since the sources are nearly 100\% identified
\citep{gio03}, we are able to present this information for three types
of sources, AGNs, groups and clusters of galaxies and stars.

In Figure 10 we show the histogram of the two X-ray colors, HR1 and
HR2. On average, clusters tend to be harder than AGNs, which are, on
average, harder than stars. Median values of (HR1, HR2) are (0.71, 0.20), 
(0.45, 0.14), and (0.18, -0.02) for clusters, AGNs, and stars
respectively. The HR2 versus HR1 scatter plot in Figure 11 shows that
the three classes of sources may be distinguished to some extent by
their X-ray colors.

Figure 12 shows the measured extent histogram for sources with
non-zero extent likelihood. As expected, many clusters are found to be
extended and most stars and AGN have small extents
($<$25\arcsec). Determining extent is substantially more difficult
than that of existence, and it hindered both by poor statistics as
well as very good statistics as discussed in Section 3. Imposing a
higher threshold on the extent likelihood does not ameliorate this
problem, while decreasing the sample size. For example only 20 of the
42 sources with extent likelihoods $\geq10$ are groups, clusters, or
an isolated galaxy.  The rest are likely point sources (AGNs, BL Lacs,
or stars). Thus the extent determination must be used cautiously.

\subsection{Finding Clusters Using their X-ray Properties}

Fourteen percent of the NEP sample are clusters, a typical fraction of
high - latitude X-ray sources at these flux levels. Much effort has
been expended over the past fifteen years compiling samples of X-ray
selected clusters of galaxies. This work involves sifting the clusters
out of the much larger total sample. Statistically, clusters are hard
extended X-ray sources, which may provide a way of isolating them
using the X-ray data, i.e. without acquiring any new data.

We may use our data to assess how effective this method is for the
RASS. Figure 13 shows that clusters preferentially occupy the upper
right portion of a HR1 - extent scatter plot. Quantitatively,
considering only sources with $\geq 100$ net photons in order to
improve the statistics of these two quantities, 72\% of all clusters
are recovered cutting at HR1 $\geq0.5$ and extent $\geq25\arcsec$,
which is also a 28\% false negative rate. But 22\% of all sources with
these cuts are AGN or stars, or false positives. So the RASS X-ray
data alone are only of marginal usefulness for finding clusters at the
depth of the NEP survey. Selection of clusters based on RASS data
alone might be more useful for the average RASS exposure, which is
lower than that at the NEP, where most clusters have larger apparent
sizes since they are not as distant. However this assertion needs
confirmation by a more careful analysis.

\section{SUMMARY AND CONCLUSIONS}

In this paper we presented a catalog containing the X-ray properties
and the identification content of the sources in the {\it ROSAT} North
Ecliptic Pole Survey. We described the selection criteria and the
solid angle surveyed as a function of count rate and give the
conversion to flux. We gave an example of the three major classes of
X-ray sources in the survey and then described the average X-ray
properties of these three classes.

The {\it ROSAT} NEP survey is unique. It is a complete flux limited
sample of X-ray sources above a limit of $\sim 2 \times 10^{-14}$ erg
cm$^{-2}$ s$^{-1}$ in the 0.5 - 2.0 keV band. This depth is comparable
to all but the deepest {\it ROSAT} pencil beam surveys.  Further, the
survey area is contained in a contiguous region of 80.6 square
degrees, comparable to all but the largest solid angle X-ray
surveys. This combination of parameters yields the deepest observation
of the X-ray sky over such a large contiguous solid angle completed so
far. Lastly, we have identified all but 2 of the 442 sources in the
survey, both of which may be statistical fluctuations.  The survey
comprises 219 AGNs, 149 stars, 62 clusters of galaxies (recalling that
RX J1724.1+7000 and RX J1724.2+6956 are two pieces of the same
object), 8 BL Lacs, one possibly interacting galaxy and one planetary
nebula.

\acknowledgments

The joint Hawaii-MPE project reported here originated with a
conversation between Pat Henry and Joachim Tr{\"u}mper at the May 1988
{\it ROSAT} International Users Committee meeting in Leicester. Many
thanks go to the entire {\it ROSAT} team, particularly Joachim
Tr{\"u}mper. F. Boese kindly provided the digital data for the RASS
point spread function, S. Snowden clarified at what point in the
analysis chain the various components making up the total effective
area were incorporated and G{\"u}nther Hasinger helped our
understanding of various aspects of the RASS. We are very grateful to
our sponsors, without whom this long program could not have been
completed. Support came from the US National Science Foundation (AST
91-19216 and AST 95-00515), the US National Aeronautics and Space
Administration (NGT5-50175, GO-5402.01-93A and GO-05987.02-94A), the
North Atlantic Treaty Organization (CRG91-0415), the Achievement
Rewards for College Scientists Foundation, the Smithsonian
Institution, the Italian Space Agency ASI-CNR, the Bundesministerium
f{\"u}r Forschung (BMBF/DLR) and the Max-Planck-Gesellschaft
(MPG). This paper was written at the MPE with the generous support of
the Alexander von Humboldt Foundation.

\clearpage

\begin{figure}
\plotone{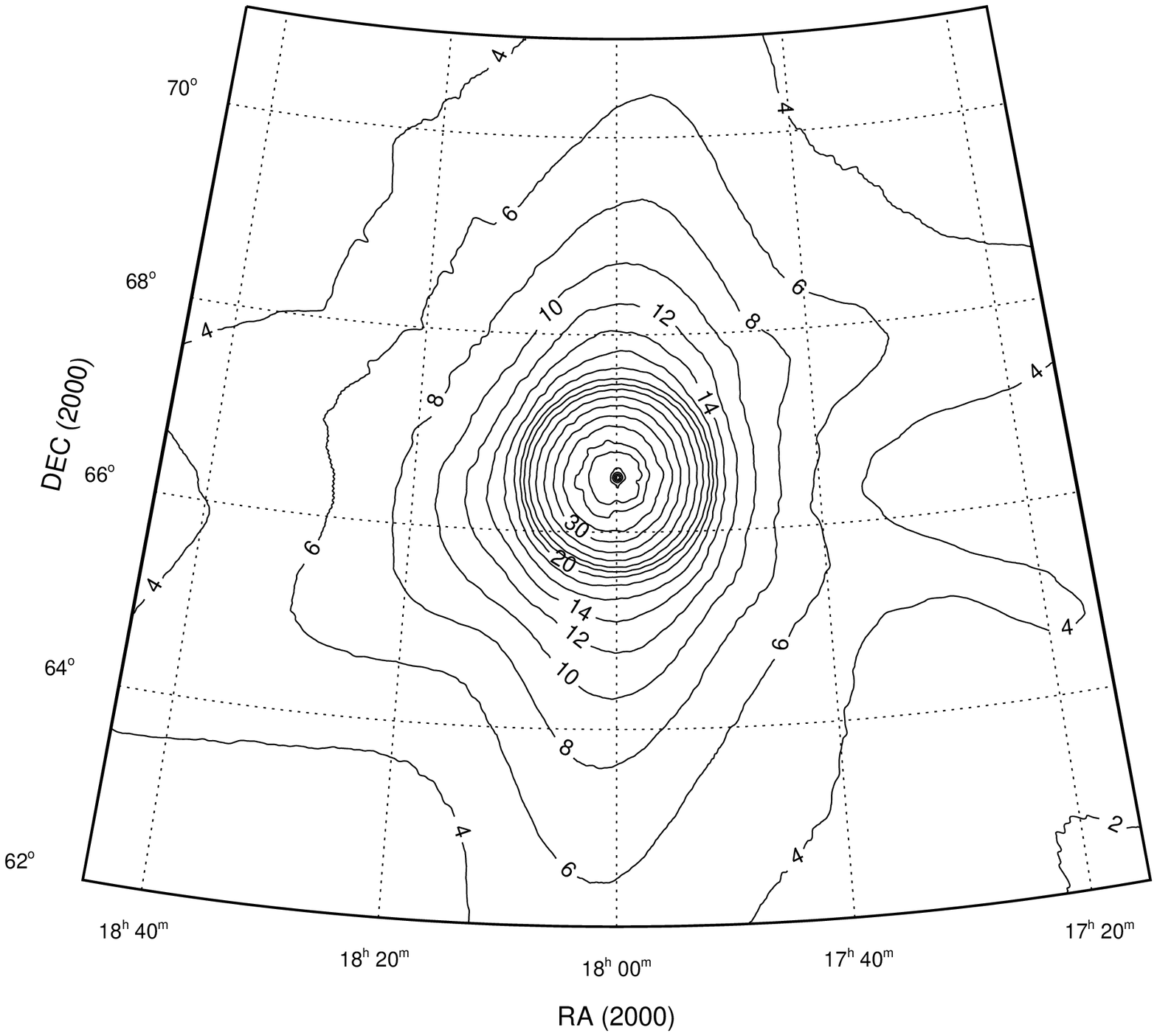}
\caption{{\it ROSAT} NEP Survey exposure in kiloseconds. The
equivalent on - axis exposure, including the effects of telescope
vignetting and window support structure both normalized to unity on
axis, is plotted. This figure originally appeared in \citet{vog01} and
\citet{mul011}; we include it here for completeness. \label{fig1}}
\end{figure}

\begin{figure}
\plotone{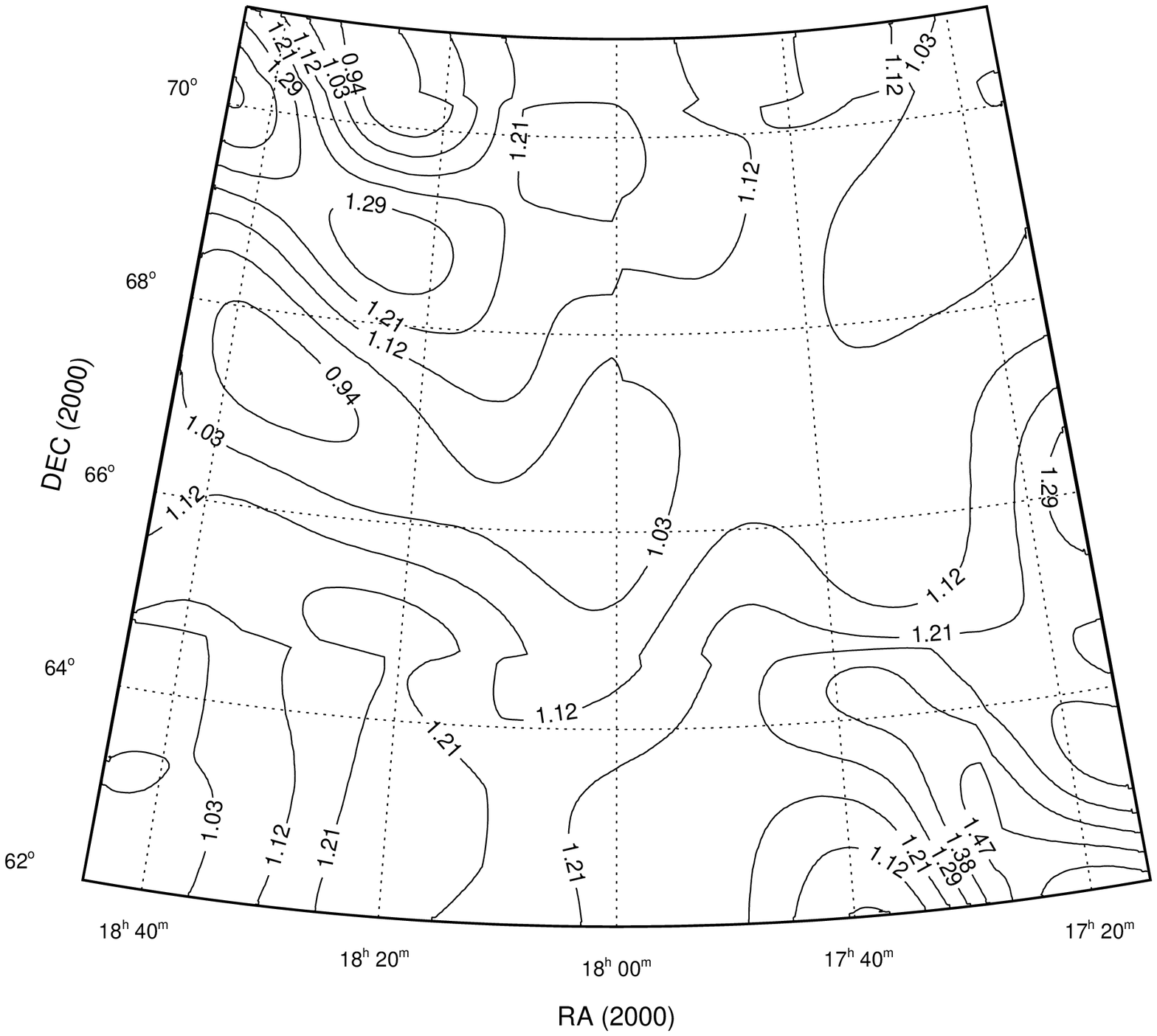}
\caption{{\it ROSAT} NEP Survey background rate in 1 $\times~10^{-3}$
cts s$^{-1}$ arcmin$^{-2}$ in the 0.1 - 2.4 keV (PHA channels 11 -
235) band. The minimum, median, and maximum background rate for the
NEP X-ray sources are (0.9, 1.1, 1.6) in the same units,
respectively. The minor discontinuities in some of the contours are at
the boundaries of the 6\fdg4 $\times$ 6\fdg4 sky fields within which
the background was independently determined. This figure originally
appeared in \citet{mul011}; we include it here for
completeness. \label{fig2}}
\end{figure}

\begin{figure}
\plotone{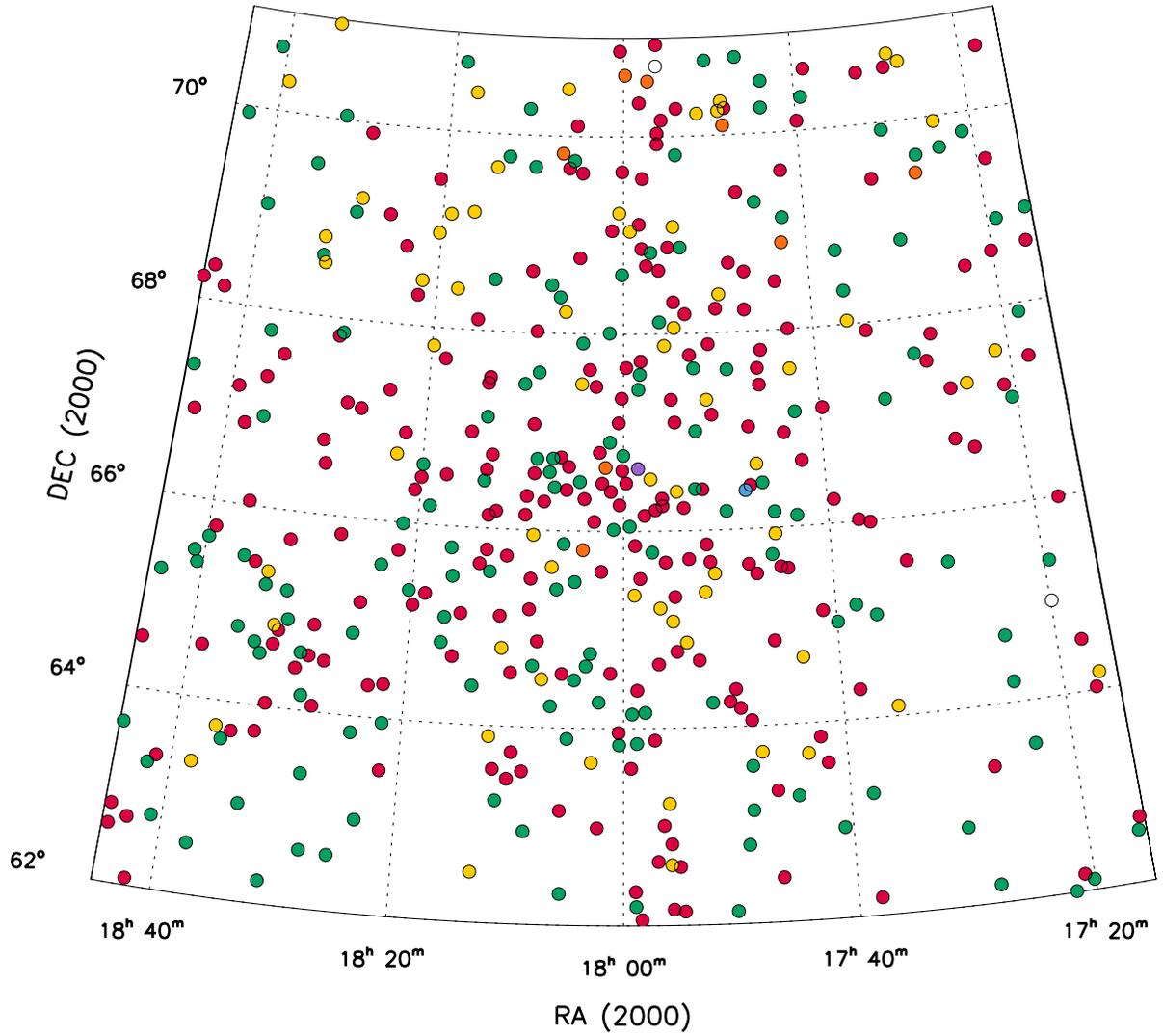}
\caption{Identification content and distribution of sources in the 
{\it ROSAT} NEP Survey. The color coding of the identifications is
as follows: AGN: red, star: green, group or cluster: yellow,
BL Lac: orange, galaxy: blue, planetary nebula: purple, no ID: white.
\label{fig3}}
\end{figure}

\begin{figure}
\plotone{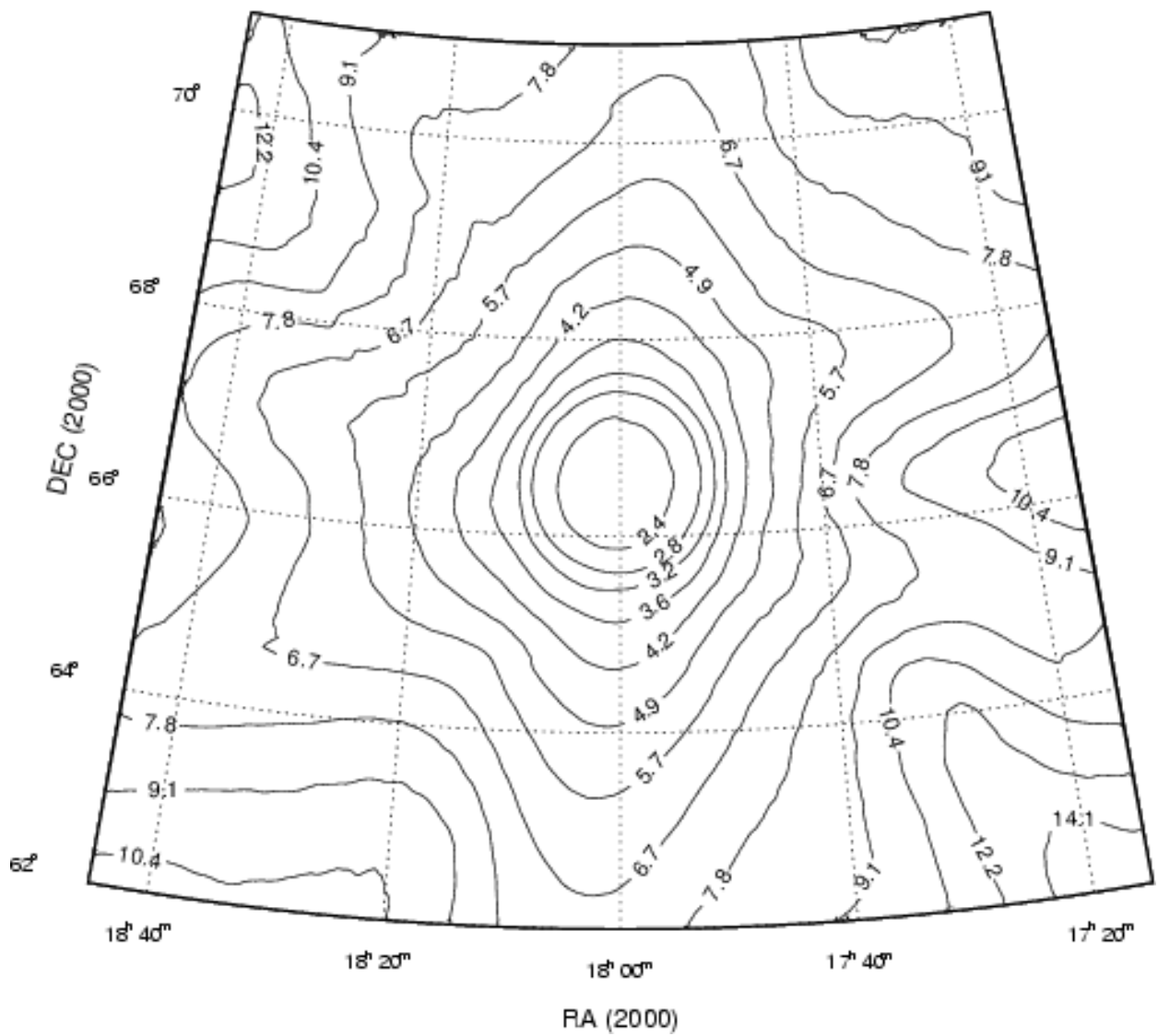}
\caption{{\it ROSAT} NEP Survey selection function in 1
$\times~10^{-3}$ cts s$^{-1}$ in the 0.1 - 2.4 keV band and in the
5\arcmin ~radius flux aperture. The contours are the minimum counting
rate a source must have to meet the 4$\sigma$ flux measurement
criterion. To convert to unabsorbed flux in the 0.5-2.0 keV band in
units of erg cm$^{-2}$ s$^{-1}$, multiply by 1.02 $\times~10^{-11}$
for extragalactic point source total flux, 6.25 $\times~10^{-12}$ for
galactic source total flux, or 1.08 $\times~10^{-11}$ for cluster flux
within a 5\arcmin ~radius flux aperture. \label{fig4}}
\end{figure}

\begin{figure}
\plotone{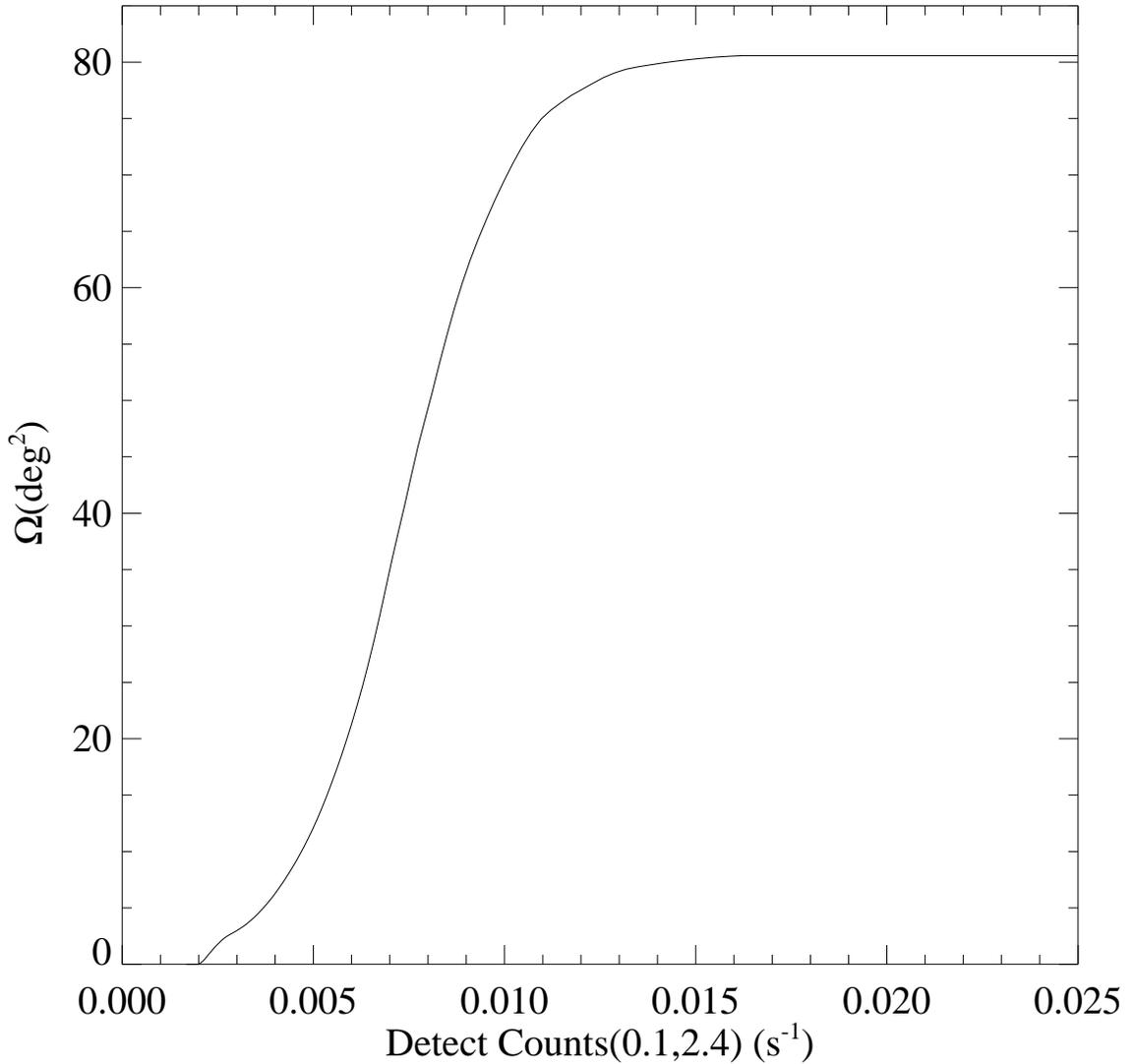}
\caption{The total solid angle observed during the {\it ROSAT} NEP
Survey as a function of cts s$^{-1}$ in the 0.1 - 2.4 keV band and in
the 5\arcmin ~radius flux aperture. To convert to unabsorbed flux in
the 0.5-2.0 keV band in units of erg cm$^{-2}$ s$^{-1}$, multiply by
1.02 $\times~10^{-11}$ for extragalactic point source total flux, 6.25
$\times~10^{-12}$ for galactic source total flux, or 1.08
$\times~10^{-11}$ for cluster flux within a 5\arcmin ~radius flux
aperture. \label{fig5}}
\end{figure}

\begin{figure}
\plotone{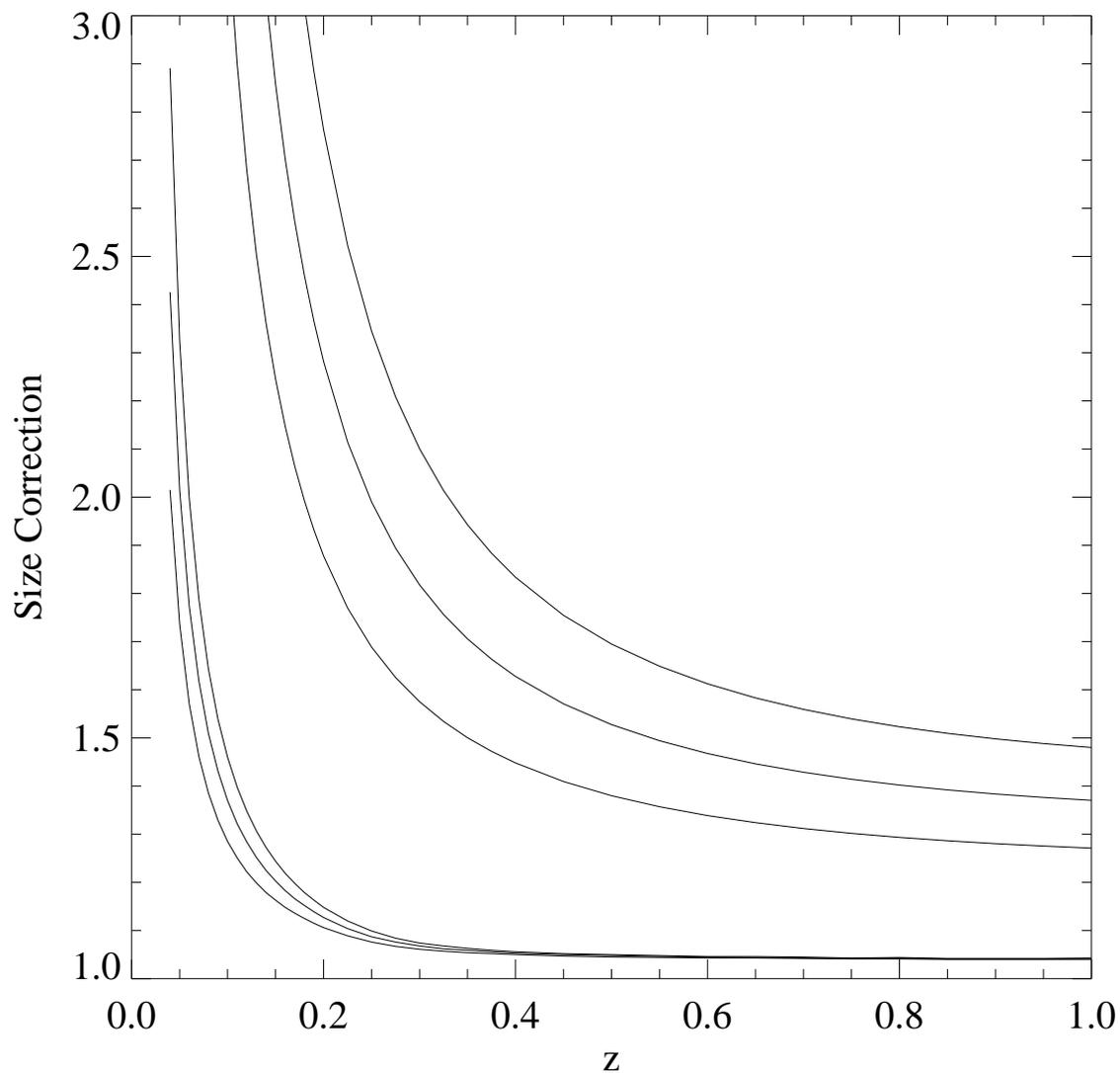}
\caption{Size correction for the {\it ROSAT} NEP Survey (lower curves)
and Extended Medium Sensitivity Survey (upper curves) clusters as a
function of redshift. For each survey the center line assumes a
non-evolving core radius of 180 h$^{-1}_{70}$~kpc, while the lower and
upper lines assume 140 or 220 h$^{-1}_{70}$~kpc respectively. The NEP
curves assume a cluster temperature of 3 keV, the median iterated
value. \label{fig6}}
\end{figure}

\begin{figure}
\plotone{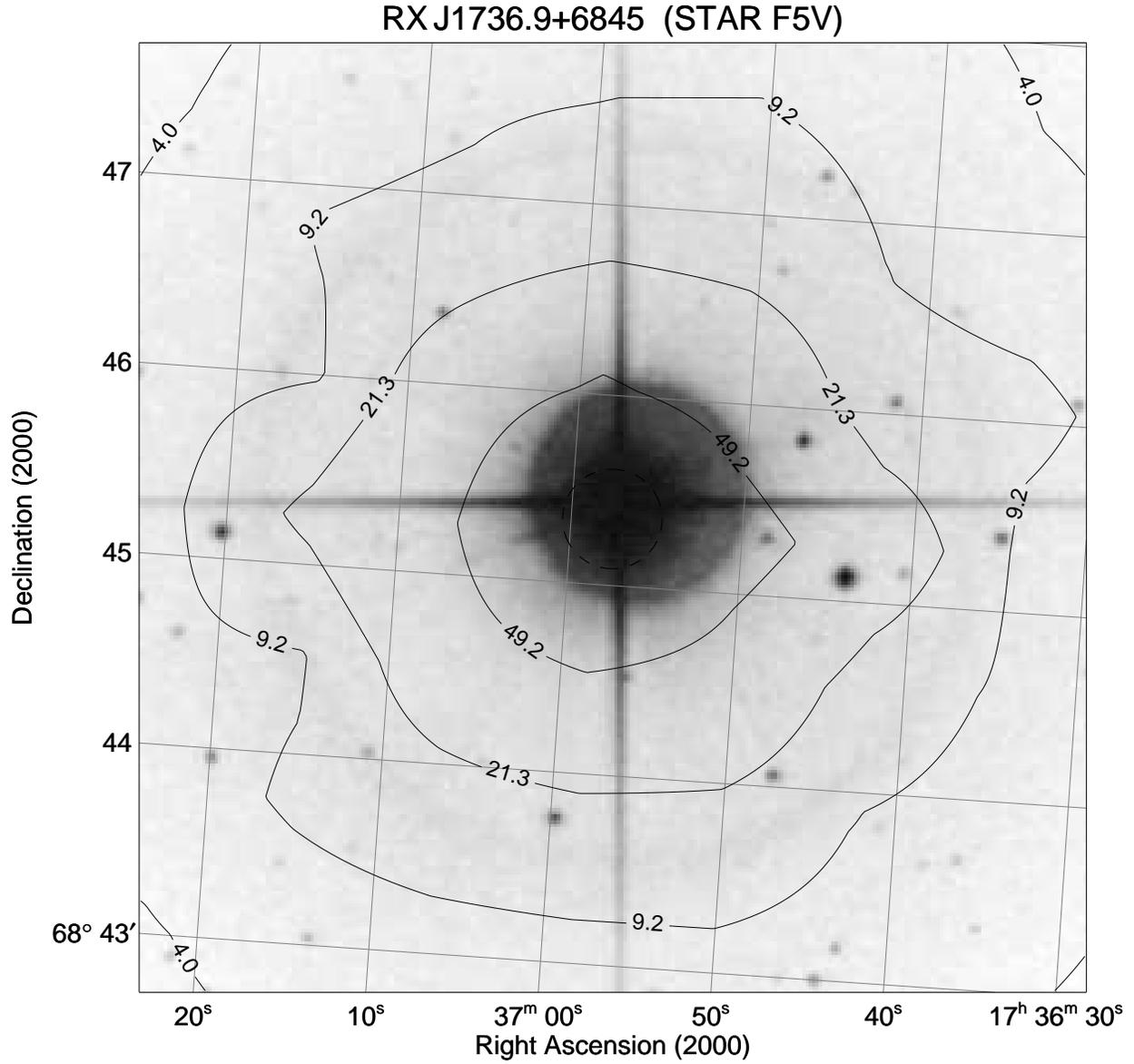}
\caption{RX J1736.9+6845, the V = 4.9 F5V star $\Omega$ Dra
(SAO17576). The optical image is from the POSS-II red plates of the
Digitized Sky Survey.  The contours are an adaptively smoothed RASS
photon image in the broad (0.1-2.4 keV) band that has
45\arcsec~$\times$ 45\arcsec~pixels. Background has not been
subtracted. Contours are labeled in units of 10$^{-3}$ counts s$^{-1}$
arcmin$^{-2}$. This figure originally appeared in \citet{mul011}.
\label{fig7}}
\end{figure}

\begin{figure}
\plotone{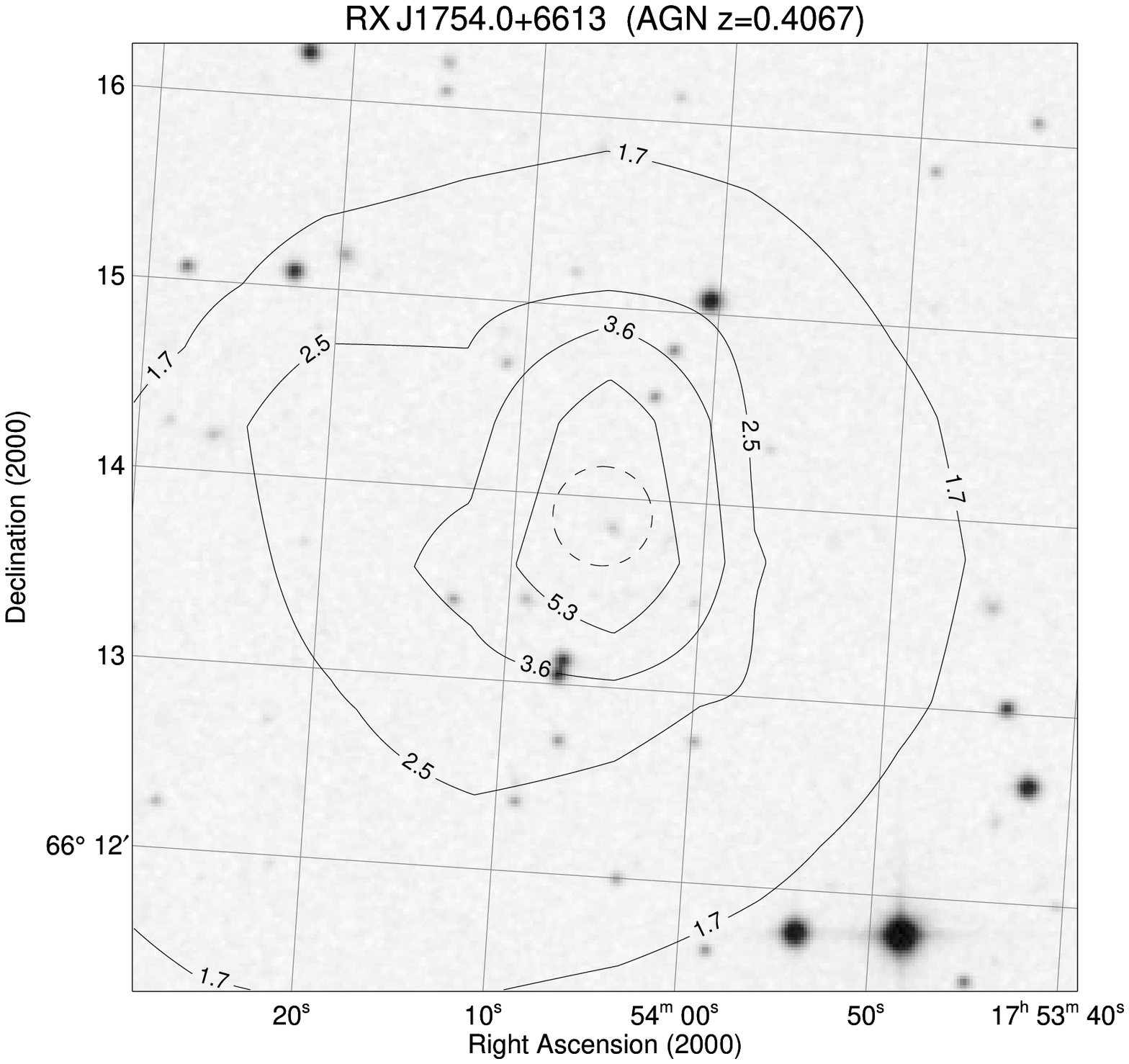}
\caption{RX J1754.0+6613, a type 1 AGN at redshift 0.4067.  The
optical image is from the POSS-II red plates of the Digitized Sky
Survey.  The contours are an adaptively smoothed RASS photon image in
the broad (0.1-2.4 keV) band that has 45\arcsec ~$\times$ 45\arcsec~
pixels. Background has not been subtracted. Contours are labeled in
units of 10$^{-3}$ counts s$^{-1}$ arcmin$^{-2}$. The dashed circle
shows the 90\% confidence position error circle (15\farcs7 radius) in
which the faint (O = 19.5, O - E = 0.64) AGN is visible. This figure
originally appeared in \citet{mul011}. \label{fig8}}
\end{figure}

\begin{figure}
\plotone{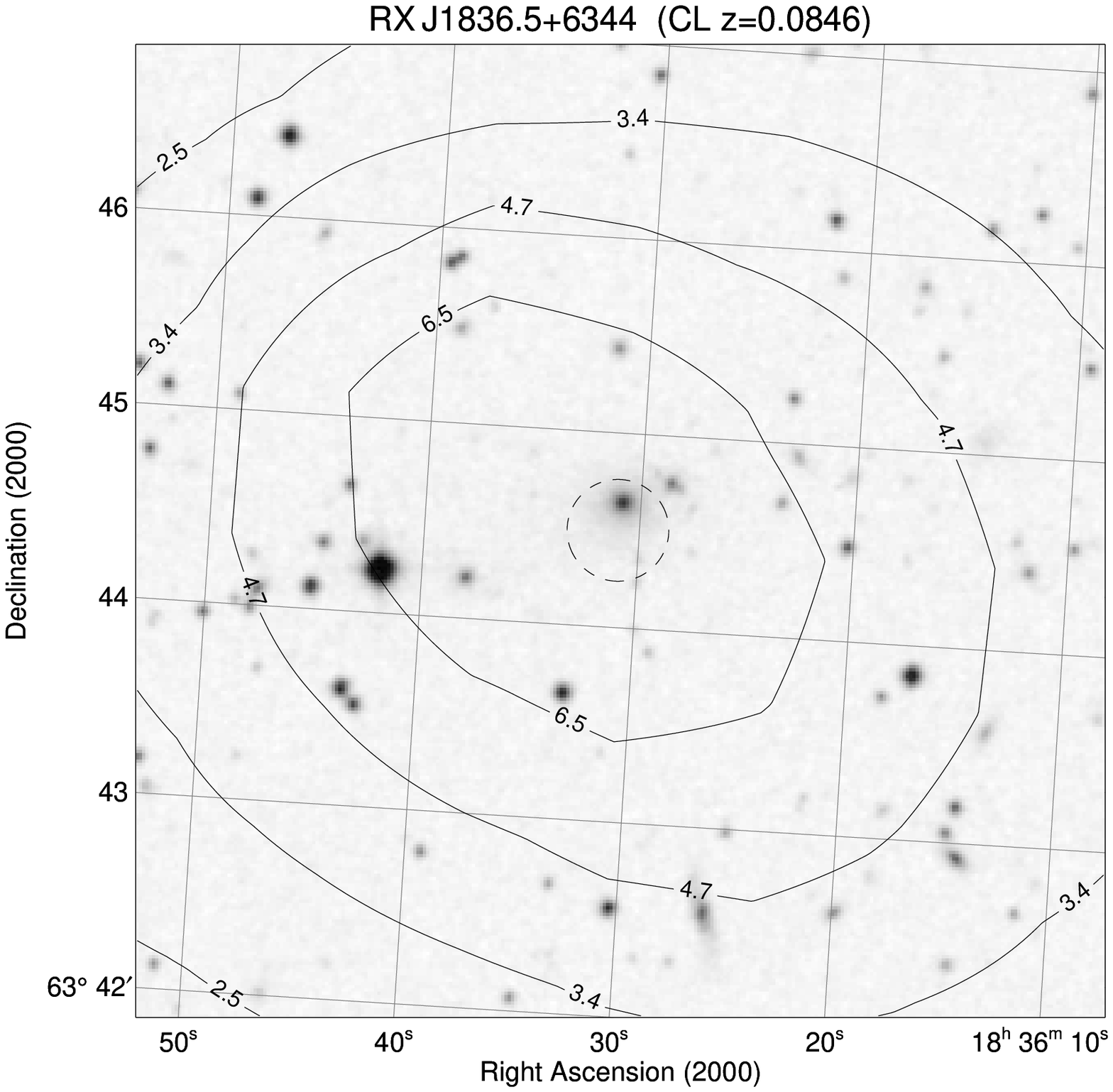}
\caption{RX J1836.5+6344, a cluster at redshift 0.0846.  The optical
image is from the POSS-II red plates of the Digitized Sky Survey.  The
contours are an adaptively smoothed RASS photon image in the broad
(0.1-2.4 keV) band that has 45\arcsec ~$\times$ 45\arcsec~
pixels. Background has not been subtracted. Contours are labeled in
units of 10$^{-3}$ counts s$^{-1}$ arcmin$^{-2}$. The dashed circle
shows the 90\% confidence position error circle (15\farcs7 radius) in
which the brightest cluster galaxy (O = 15.8, O - E = 2.24) is
visible. This figure originally appeared in \citet{mul011}. \label{fig9}}
\end{figure}

\begin{figure}
\epsscale{.80}
\plottwo{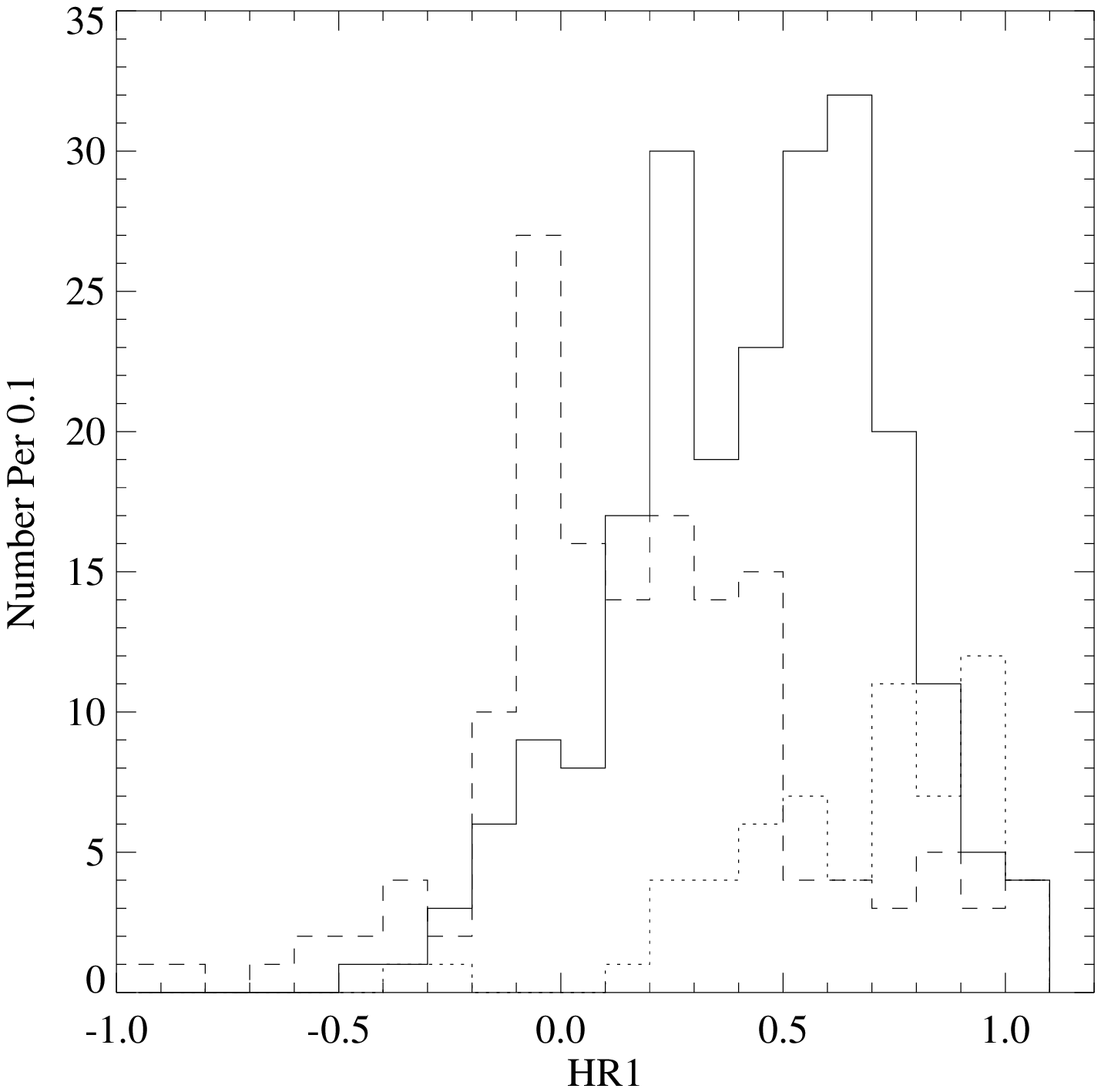}{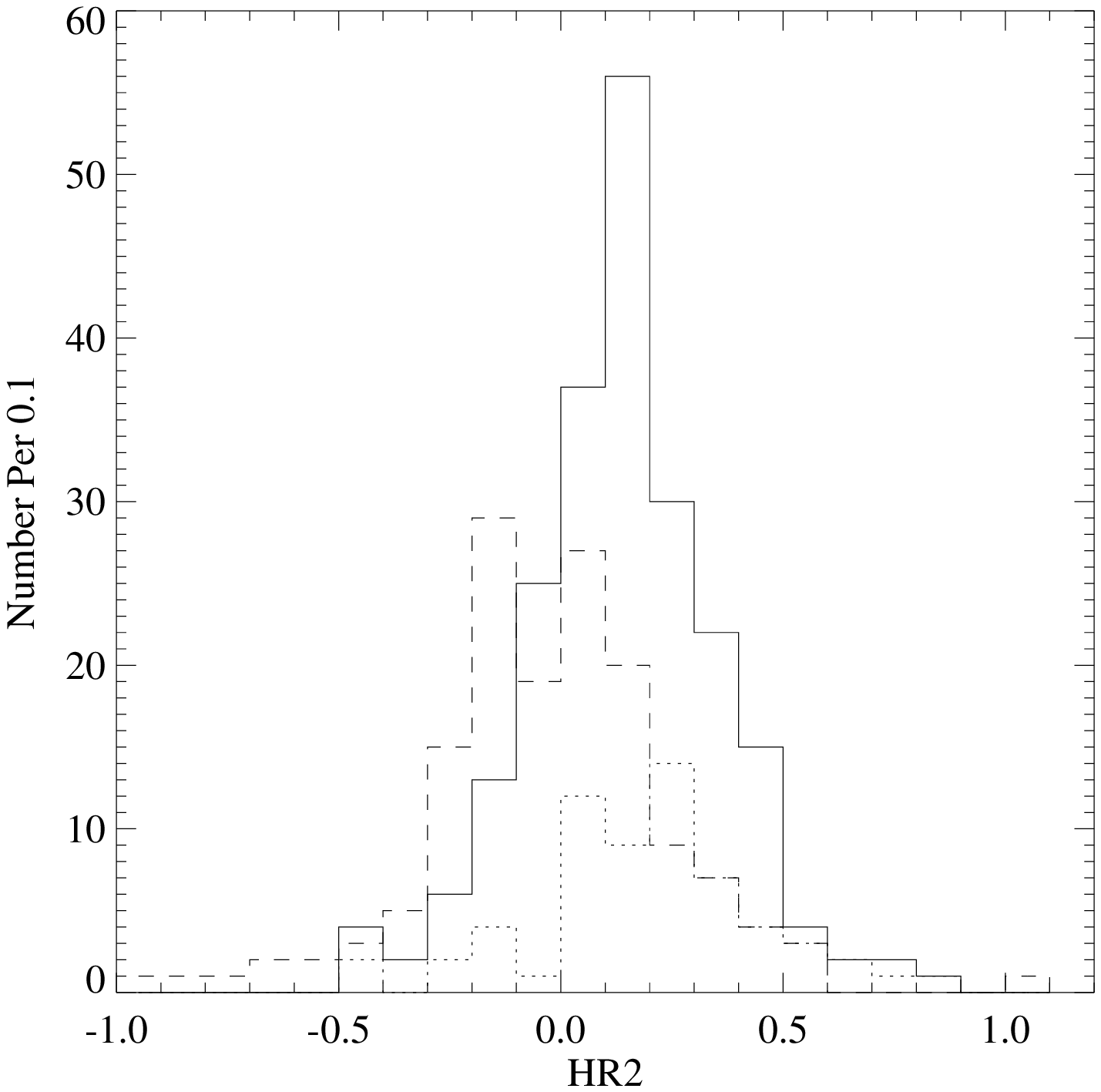}
\caption{Spectral hardness ratios. AGNs are solid, stars dashed and
groups and galaxy clusters dotted. Panel a (left) shows HR1; panel
b (right) shows HR2. The rightmost bin contains only objects with
hardness ratio exactly 1.0.\label{fig10}}
\end{figure}

\begin{figure}
\plotone{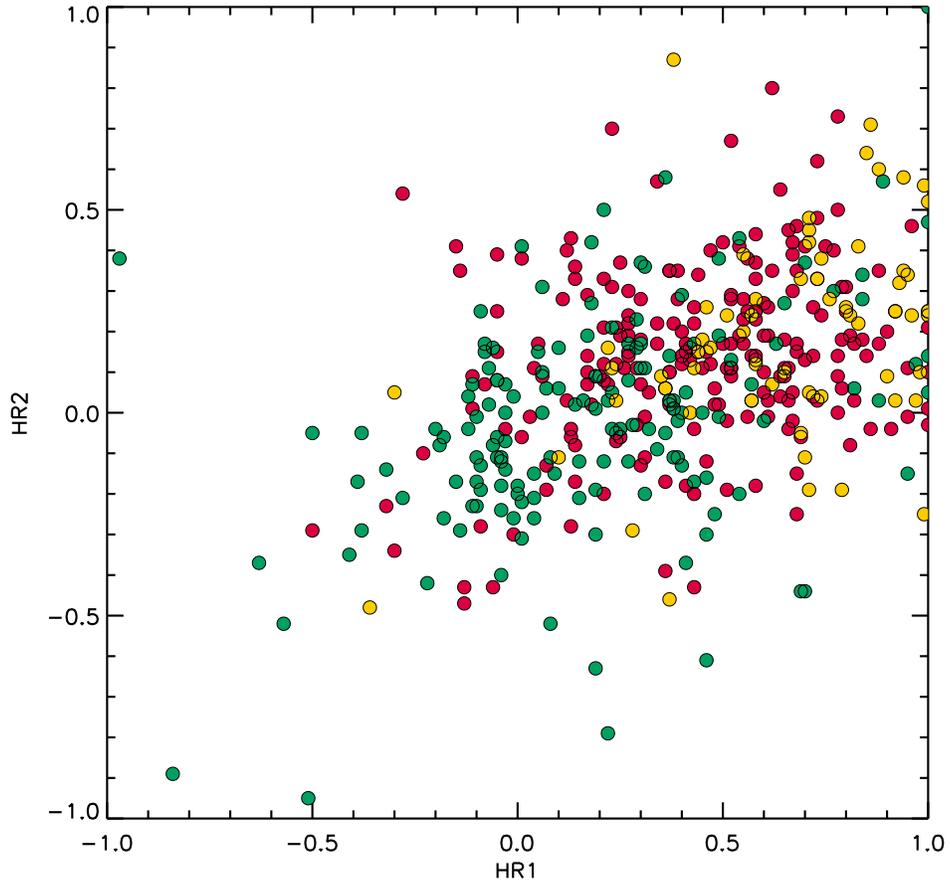}
\caption{HR2 versus HR1 scatter plot. AGNs are red, stars green, and
groups and galaxy clusters yellow. These two quantities provide some
discrimination among the three classes. \label{fig11}}
\end{figure}

\begin{figure}
\plotone{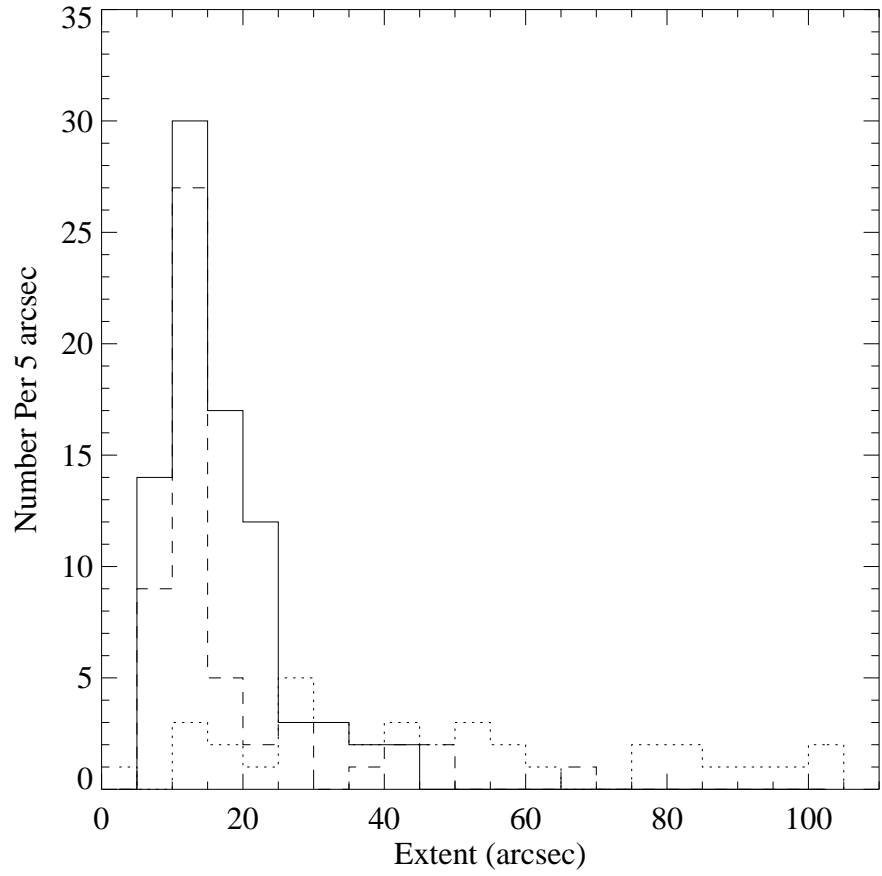}
\caption{Measured angular extent for NEP sources with non-zero extent
likelihood. AGNs are solid, stars dashed and groups and galaxy
clusters dotted. \label{fig12}}
\end{figure}

\begin{figure}
\plotone{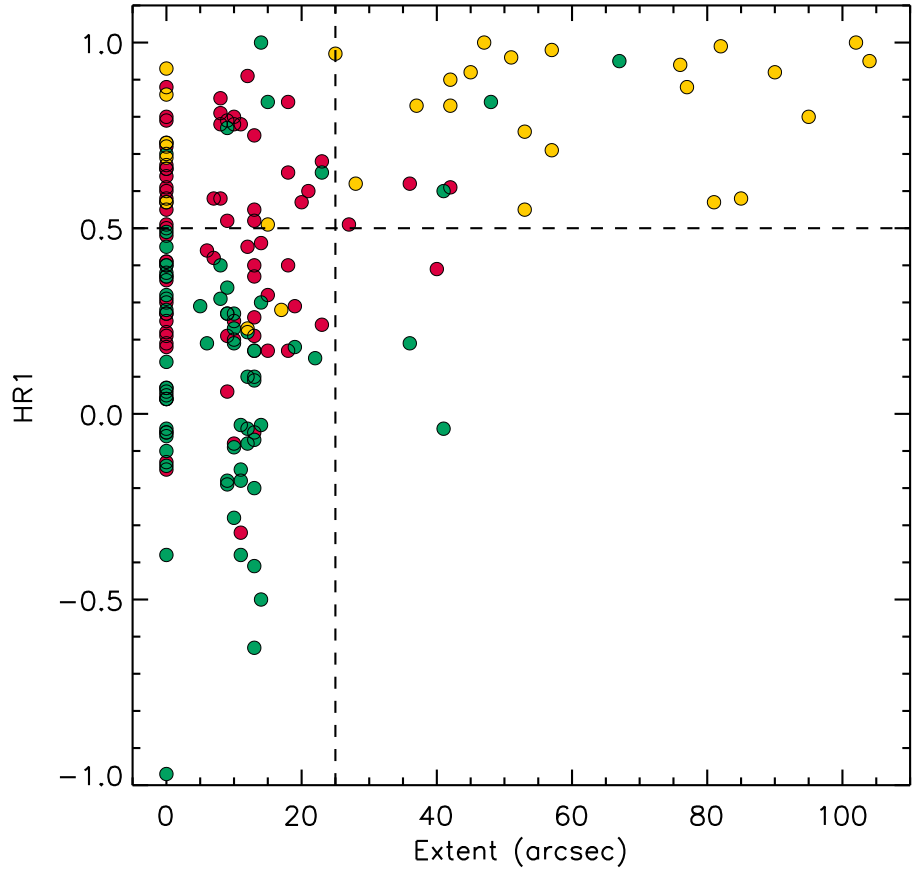}
\caption{HR1 versus X-ray extent for the NEP Survey sources with
$\ge$~100 net photons. AGN are red, stars are green, and groups and
clusters of galaxies are yellow. The dashed lines show the extent and
hardness ratio cuts discussed in the text. \label{fig13}}
\end{figure}

\clearpage
\begin{deluxetable}{ccccc}
\tabletypesize{\footnotesize}
\tablewidth{0pt}
\tablecaption{Selection Criteria for the {\it ROSAT} NEP X-ray Survey\label{tbl-1}}
\tablehead{
\colhead{Right Ascension}        & \colhead{Declination} & 
\colhead{Energy Band}            & \colhead{Exist L}     & 
\colhead{Detect Count Rate S/N}  
}
\startdata
$17^{h}15^{m}<\alpha(2000)<18^{h}45^{m}$&$62\degr<\delta(2000)<71\degr$&0.1 - 2.4 keV&$\geq$10&$>$4\\
\enddata
\end{deluxetable}

\clearpage
\begin{deluxetable}{ccccc}
\tabletypesize{\footnotesize}
\tablewidth{0pt}
\tablecaption{{\it ROSAT} North Ecliptic Pole Sky Coverage\label{tbl-2}}
\tablehead{
\colhead{Count Rate}              & \colhead{Total Gal Point Flux}      & 
\colhead{Total Ex Gal Point Flux} & \colhead{Detect Ex Gal Diffuse Flux}& 
\colhead{Solid Angle}                  
\\
\colhead{(0.1-2.4 ct s$^{-1}$)} & \colhead{(0.5-2.0 erg cm$^{-2}$ s$^{-1}$)}& 
\colhead{(0.5-2.0 erg cm$^{-2}$ s$^{-1}$)}& 
\colhead{(0.5-2.0 erg cm$^{-2}$ s$^{-1}$)}& \colhead{(deg$^{2}$)}
}
\startdata
2.02$\times$~10$^{-3}$&1.26$\times$~10$^{-14}$&2.06$\times$~10$^{-14}$&
2.18$\times$~10$^{-14}$&0.038\\
2.12$\times$~10$^{-3}$&1.32$\times$~10$^{-14}$&2.16$\times$~10$^{-14}$&
2.29$\times$~10$^{-14}$&0.298\\
3.00$\times$~10$^{-3}$&1.87$\times$~10$^{-14}$&3.06$\times$~10$^{-14}$&
3.24$\times$~10$^{-14}$&2.982\\
4.05$\times$~10$^{-3}$&2.53$\times$~10$^{-14}$&4.13$\times$~10$^{-14}$&
4.37$\times$~10$^{-14}$&6.505\\
5.10$\times$~10$^{-3}$&3.19$\times$~10$^{-14}$&5.21$\times$~10$^{-14}$&
5.51$\times$~10$^{-14}$&12.89\\
6.00$\times$~10$^{-3}$&3.75$\times$~10$^{-14}$&6.12$\times$~10$^{-14}$&
6.48$\times$~10$^{-14}$&21.32\\
7.06$\times$~10$^{-3}$&4.41$\times$~10$^{-14}$&7.20$\times$~10$^{-14}$&
7.26$\times$~10$^{-14}$&35.88\\
7.39$\times$~10$^{-3}$&4.62$\times$~10$^{-14}$&7.54$\times$~10$^{-14}$&
7.98$\times$~10$^{-14}$&40.76\\
\enddata
\end{deluxetable}

\clearpage
\begin{deluxetable}{ccccccccc}
\tabletypesize{\footnotesize}
\tablewidth{0pt}
\tablecaption{Properties of Three NEP X-ray Sources\label{tbl-3}}
\tablehead{
\colhead{RX J}        & \colhead{NEP No.}    & 
\colhead{Class}       & \colhead{Count Rate} & 
\colhead{HR1}         & \colhead{HR2}        & 
\colhead{Extent}      & \colhead{Extent L}   & 
\colhead{Exist L} 
\\
\colhead{}            & \colhead{}              & 
\colhead{}            & \colhead{(ct s$^{-1}$)} & 
\colhead{}            & \colhead{}              & 
\colhead{(\arcsec)}   & \colhead{}              & 
\colhead{} 
}
\startdata
1736.9+6845&2130&STAR&0.4655$\pm$0.0103&$-0.28\pm0.02$&$-0.21\pm0.04$&10&10&999\\
1754.0+6613&60&AGN1&0.0152$\pm$0.0010&0.42$\pm$0.07&0.16$\pm$0.07&7&1&307\\
1836.5+6344&6190&CL&0.1267$\pm$0.0065&0.92$\pm$0.05&0.03$\pm$0.06&90&265&289\\
\enddata
\end{deluxetable}

\clearpage
\begin{deluxetable}{cccccccccccc}
\tabletypesize{\tiny}
\tablewidth{0pt}
\tablecaption{{\it ROSAT} NEP X-ray Source Catalog\label{tbl-4}}
\tablehead{
\colhead{RX J}                & \colhead{$\alpha$(J2000)}       & 
\colhead{Count Rate}          & \colhead{Exposure(s)}           &
\colhead{}                    & \colhead{}                      &
\colhead{Extent(\arcsec)}     & \colhead{Exist L}               & 
\colhead{$f_{Tot}(10^{-14})$} & \colhead{ID}                    & 
\colhead{kT$/\Gamma$}
\\
\colhead{NEP No.}             & \colhead{$\delta$(J2000)}       & 
\colhead{(ct s$^{-1}$)}       & \colhead{$N_H(10^{20}cm^{-2}$)} & 
\colhead{HR1}                 & \colhead{HR2}                   &
\colhead{Extent L}            & \colhead{Map L}                 &  
\colhead{$L_{X}(10^{44}$)}    & \colhead{z}                     & 
\colhead{SC}
\\
\colhead{(1)}           & \colhead{(2)}        &   
\colhead{(3)}           & \colhead{(4)}        &   
\colhead{(5)}           & \colhead{(6)}        &   
\colhead{(7)}           & \colhead{(8)}        &   
\colhead{(9)}           & \colhead{(10)}       &   
\colhead{(11)}          
}
\input{tab4}

\end{deluxetable}

\clearpage
\begin{deluxetable}{cccccccccccc}
\tabletypesize{\tiny}
\tablewidth{0pt}
\tablecaption{Properties of Sources Not Meeting the Selection Criteria
in the Broad Band\label{tbl-5}}
\tablehead{
\colhead{RX J}                & \colhead{$\alpha$(J2000)}       & 
\colhead{Count Rate}          & \colhead{Exposure(s)}           &
\colhead{}                    & \colhead{}                      &
\colhead{Extent(\arcsec)}     & \colhead{Exist L}               & 
\colhead{$f_{Tot}(10^{-14})$} & \colhead{ID}                    & 
\colhead{kT$/\Gamma$}
\\
\colhead{NEP No.}             & \colhead{$\delta$(J2000)}       & 
\colhead{(ct s$^{-1}$)}       & \colhead{$N_H(10^{20}cm^{-2}$)} & 
\colhead{HR1}                 & \colhead{HR2}                   &
\colhead{Extent L}            & \colhead{Map L}                 &  
\colhead{$L_{X}(10^{44}$)}    & \colhead{z}                     & 
\colhead{SC}
\\
\colhead{(1)}           & \colhead{(2)}        &   
\colhead{(3)}           & \colhead{(4)}        &   
\colhead{(5)}           & \colhead{(6)}        &   
\colhead{(7)}           & \colhead{(8)}        &   
\colhead{(9)}           & \colhead{(10)}       &   
\colhead{(11)}          
}
\startdata
1738.0+6653&17 38  1.3&0.0035& 5201&$ 1.00$&$ 0.00$&  0&  4&  2.18&STAR      & 0.9& \\
2132&+66 53 36 &$\pm0.0017$&$\ldots$&$\pm0.09$&$\pm0.00$&  0& 11&$\ldots$&$\ldots$&1.050& \\
1752.7+6804&17 52 47.0&0.0023&11323&$ 0.62$&$ 0.47$& 41&  7&  1.46&STAR      & 0.9& \\
  31&+68  4 48 &$\pm0.0009$&$\ldots$&$\pm0.25$&$\pm0.28$&  1&  8&$\ldots$&$\ldots$&1.050& \\
1755.3+6416&17 55 21.7&0.0030& 8810&$ 1.00$&$-0.06$& 73&  7&  3.11&CL        & 2.0& \\
3391&+64 16 39 &$\pm0.0012$&3.31&$\pm0.17$&$\pm0.26$& 18&  8&  0.08&0.2837&1.062& \\
\enddata
\end{deluxetable}

\end{document}

%% file: tab4.tex
\startdata
1715.4+6239&17 15 25.3&0.0173& 2062&$-0.09$&$-0.28$& 15& 27& 13.99&AGN2      & 2.0& \\
1239&+62 39 34 &$\pm0.0038$&2.59&$\pm0.21$&$\pm0.31$&  1& 14&  4.88&0.8500&1.050& \\
1715.6+6856&17 15 41.4&0.0148& 3474&$ 0.43$&$ 0.17$&  0& 45&  9.24&STAR      & 0.9& \\
1240&+68 56 31 &$\pm0.0028$&$\ldots$&$\pm0.18$&$\pm0.20$&  0& 19&$\ldots$&$\ldots$&1.050& \\
1715.6+6231&17 15 41.6&0.0170& 1996&$ 0.46$&$-0.30$&  0& 27& 10.62&STAR      & 0.9& \\
1241&+62 31 33 &$\pm0.0039$&$\ldots$&$\pm0.27$&$\pm0.22$&  0& 13&$\ldots$&$\ldots$&1.050& \\
1716.2+6836&17 16 14.4&0.1192& 3935&$ 0.72$&$ 0.26$&  0&972&115.39&AGN1      & 2.0& \\
1270&+68 36 36 &$\pm0.0059$&3.67&$\pm0.04$&$\pm0.05$&  0&497& 32.28&0.7770&1.050& \\
\enddata